\documentclass{aa}

\usepackage{natbib}
\usepackage{graphicx}
\usepackage{txfonts}
\usepackage{subfigure}

\bibpunct{(}{)}{;}{a}{}{,}
\begin{document}

\title{A multi-particle model of the \object{3C 48} host}

\author{J. Scharw\"achter\inst{1} \and
           A. Eckart\inst{1} \and
           S. Pfalzner\inst{1} \and
           J. Zuther\inst{1} \and
           M. Krips\inst{1,2} \and
           C. Straubmeier\inst{1}
}

\offprints{J. Scharw\"achter, \email{scharw@ph1.uni-koeln.de}}

\institute{I. Physikalisches Institut, Universit\"at zu K\"oln, Z\"ulpicher Str. 77, 50937 K\"oln \and
IRAM, 300, rue de la Piscine, Domaine Universitaire, 38406 Saint Martin d'H\'eres}

\date{Received (date); accepted (date)}

\abstract{The first successful multi-particle model for the host of the
well-known quasi-stellar object (QSO) \object{3C 48} is reported. 
It shows that the 
morphology and the stellar velocity field of the \object{3C 48} host can be 
reproduced by the merger of two disk galaxies. The conditions of the 
interaction are similar to those used
for interpreting the appearance of the ''Antennae'' (\object{NGC 4038/39})
but seen from a different viewing
angle. The model supports the controversial hypothesis that \object{3C 48A}
is the second nucleus of a merging galaxy, and it suggests a 
simple solution
for the problem of the missing counter tidal tail.

}

\maketitle


\section{Introduction\label{sec:intro}}

\object{3C 48} \citep{2001AJ....121.2843B}
was the first QSO to be discovered optically
\citep{1961ST....21.148M, 1963ApJ...138...30M} and 
the first QSO to be directly identified with a host galaxy 
\citep{1982Natur.296..397B}. 
Its basic properties are listed in 
Table~\ref{tab:3c48data}.
\object{3C 48} has attracted much attention regarding the proposed 
evolutionary
sequence of active nuclei \citep{1988ApJ...325...74S}. 
According to this scheme,
interactions and mergers of galaxies trigger an evolution via 
ultra-luminous infrared galaxies (ULIRGs) to QSOs.
The observational evidence, however, is hampered by the fact that 
many transitionary objects show only dubious indications of past or
recent mergers.
Clarification requires detailed
multi-particle modeling which helps with disentangling the complex spatial
structure of merger remnants.
\begin{table}
        \caption{Basic properties of \object{3C 48}.}
        \label{tab:3c48data}
         $$
        \begin{array}{lll}
        \hline
        \noalign{\smallskip}
        \hline
        \mathrm{RA_{2000}} & \mathrm{01h37m41.3s} 
        & $\citet{1998AJ....116..516M}$ \\
        \mathrm{Dec_{2000}} &  \mathrm{+33d09\arcmin 35\farcs 1}
        & $\citet{1998AJ....116..516M}$\\
        \mathrm{Redshift} & 0.367 & 
        $\citet{2001AJ....121.2843B}$\\
        \mathrm{Luminosity\ Distance} & 1,581\ \mathrm{Mpc^a} & \\
        \mathrm{Angular\ Size\ Distance} & 846\ \mathrm{Mpc^a} & \\
        \mathrm{Scaling\ Factor} 
        & 1 \arcsec \approx 4.1\ \mathrm{kpc} & \\
        \noalign{\smallskip}
        \hline
        \end{array}
         $$
        \begin{list}{}{}
        \item[$^{a}$] 
          $\mathrm{H_0=75\ km\ s^{-1}\ Mpc^{-1}}$ and 
        $q_0=0.5 $ will be adopted throughout the paper.
        \end{list}
\end{table}

\object{3C 48} is an example of a transitionary object with prototypal 
properties in many respects: 
It has the typical far-infrared excess, originating from thermal radiation of 
dust which is heated by the quasar nucleus and by newly forming 
stars in the host galaxy \citep{1985ApJ...295L..27N, 1991AJ....102..488S}. 
Large amounts of molecular gas 
\citep{1993AAS...182.0624S, 1997A&A...322..427W}
indicate the possibility of a young stellar population in the host.
Finally, long-slit spectroscopy gives evidence for an ongoing starburst
in the host which currently seems to be close to its maximum 
activity \citep{2000ApJ...528..201C}.
But the merger scenario for \object{3C 48} is still unclear:
Indeed, the host has a significant tail-like extension to the 
northwest whose tidal origin is rather compelling with 
regard to the kinematics and
ages of its stars \citep{2000ApJ...528..201C}. However, the nature of the
apparent second nucleus \object{3C 48A} about $1\arcsec $ northeast of the 
QSO \citep{1991AJ....102..488S}  
and the location of the expected counter tidal tail
remain an unsolved problem. 
\object{3C 48A} could as well be due
to the radio jet \citep{1991Natur.352..313W} interacting with the 
dense interstellar medium  \citep{1999A&A...343..407C}. 
A feature at the southeast of the
\object{3C 48} host, previously interpreted as a counter tidal tail
\citep{1999MNRAS.302L..39B}, has 
turned out to be a background galaxy \citep{2000ApJ...528..201C}.
Instead, a counter tail extending from the southeast
to the southwest is suspected \citep{2000ApJ...528..201C} but not yet 
identified. \citet{2000ApJ...528..201C} suggest that such a
location of the two tidal tails might be explicable by a certain projection
of the merger scenario used to simulate the ''Antennae''.  

This paper reports the first successful multi-particle model for 
the \object{3C 48} host. Suggesting simple solutions for the \object{3C 48A}
problem and the counter tail problem, the model largely resolves doubts
about the merger hypothesis for \object{3C 48}.

\section{Methods \label{sec:methods}}

\subsection{Numerical simulations}

The stellar-dynamical 3-dimensional 
N-body simulations are performed with TREESPH 
\citep{1989ApJS...70..419H}, a tree code used in its non-collisional mode.
Stable initial particle distributions for the model 
galaxies are set up with BUILDGAL \citep[see][]{1993ApJS...86..389H}.
The spatial
orientation of the two disk galaxies is parameterized by their inclinations
$i$ with respect to the orbital plane and their pericentric arguments
$\omega $ as introduced by \citet{1972ApJ...178..623T}. Both
galaxies have extended mass distributions so that the orbit of their 
encounter is not Keplerian but decaying. Two descriptions can be
used to characterize these orbits: 
The first one is a pseudo-Keplerian description using
the parameters of eccentricity $e$, pericentric distance 
$r_{\mathrm{peri}}$, 
and angle to pericentre $\Omega_{\mathrm{peri}}$ of 
the corresponding Keplerian orbit
for which the total mass of each galaxy is associated with a point mass 
at the
respective centre of mass. The second one is a direct description of the
decaying orbit using the true apocentric and pericentric distances 
$r_{\mathrm{apo}}$ and $r_{\mathrm{peri}}$ of the 
first passage to define 
an eccentricity in its generalized formulation 
$e = (r_{\mathrm{apo}}-r_{\mathrm{peri}})/(r_{\mathrm{apo}}+r_{\mathrm{peri}})$.
For convenience, the system of  
units, which remains intrinsically scale-free, 
is scaled to the system suitable for \object{3C 48}.
 
The results of the simulations are analyzed as 2-dimensional projections.
In order to mock the pixel array data of imaging observations, 
the particles are sorted into a $512\times 512$ grid. The virtual pixel 
values are computed by adding up all particles located in 
a grid cell along the line-of-sight. 
Without any special weighting of a nuclear component, 
the mock images are comparable to 
QSO-subtracted images of the \object{3C 48} host.
Spectra for each grid cell are generated by sorting
the particles into velocity channels according to their respective 
line-of-sight velocities. Thus, an
average stellar line-of-sight velocity is assigned to each virtual pixel.
The resulting data arrays are spatially smoothed by Gaussian convolution 
and converted
into FITS format to facilitate the further data processing with standard
astronomical software.

\subsection{Observational data on \object{3C 48}}

\begin{figure}
        \centering\resizebox{\hsize}{!}{\includegraphics{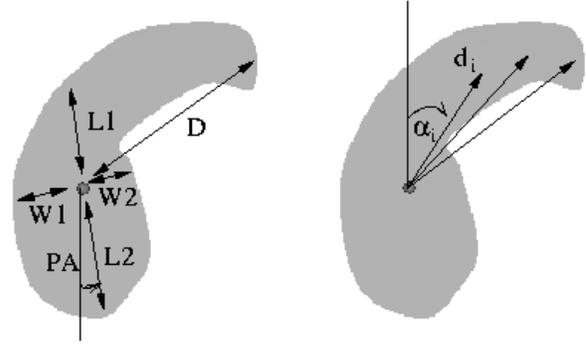}}
        \caption{Sketch of the parameters used for characterizing
        the dimensions of the \object{3C 48} host. The 
        lengths ($L1$, $L2$) are measured along the longest extension of 
        the host body (PA~$\approx 4\degr$ for the observations, 
        PA~$\approx 19\degr$ for the simulations), 
        the widths ($W1$, $W2$) are measured perpendicular to this. 
        The curvature of the tail is traced along 
        maximum intensity (right panel) by angle-distance-vectors 
        $(\alpha_i,d_i)$. }
        \label{fig:ratios}
\end{figure}

The data presented 
by \citet{2000ApJ...528..201C} are used for 
comparing the simulations with observations.
They provide information about the optical 
surface-brightness of the QSO-subtracted \object{3C 48} host 
(Fig.~1 therein) 
and about the stellar kinematics along the four slits A, B, C, G 
(Fig.~1 and Table~2 therein).
In reference to these data, the basic proportions of the main body of 
the \object{3C 48} host 
are classified by dimensionless length ratios (left panel of
Fig.~\ref{fig:ratios} and left column of Table~\ref{tab:ratios}). 
Such a comparison is independent of the length scaling, in contrast to the
comparison of line-of-sight
velocities which requires a positioning of the four slits on the mock 
image.
Having determined the final physical length unit of the simulations,
the curvature of the northwestern tidal tail is compared by 
using an angle-versus-distance plot (right panel of Fig.~\ref{fig:ratios}
and Fig.~\ref{fig:curv}).

\section{The \object{3C 48} look-alike\label{sec:results}}

Different mass ratios of the 
initial galaxies, different snapshots during the merger process, and 
different projection angles of the merger remnants were probed in a still
limited parameter study. 

The nearest \object{3C 48} look-alike is found for the merger of 
two identical galaxies whose physical and numerical properties are
given in Table~\ref{tab:galparams}. With these parameters the
galaxies are similar to spirals of type Sb. 
\begin{table}
        \caption{Initial parameters of the two identical
        galaxies in the system of units suitable for \object{3C 48}. 
        Each galaxy consists of a spherical non-rotating bulge, a
        rotating 
        exponential disk, and an isothermal halo \citep{1993ApJS...86..389H}.}
        \label{tab:galparams}
        $$
        \begin{array}{l|ccc}
        \hline
        \noalign{\smallskip}
        \hline
        \mathrm{Parameter} & \mathrm{Bulge} & \mathrm{Disk} & \mathrm{Halo} \\ 
        \hline
        \mathrm{number\ of\ particles} & 8,000 & 8,000 & 8,000 \\
        \mathrm{softening\ length\ [kpc]} & 0.21 & 0.28 
        & 1.4 \\
        \mathrm{mass\ [10^{10}M_{\sun}]} & 1.86 & 5.60
        & 32.48 \\
        \mathrm{scale\ length\ [kpc]} & 0.88 & 3.50 
        & 35.0 \\
        \mathrm{maximum\ radius\ [kpc]} & 7.0 & 52.5 
        & 105.0 \\
        \hline
        \mathrm{scale\ height\ [kpc]} &  & 0.7 & \\
        \mathrm{maximum\ height\ [kpc]} & & 7.0 & \\
        \hline
        \end{array}
        $$
\end{table}
The experimental setup is the same as used for simulations of the 
''Antennae'' -- i.e. both galaxies are symmetrically oriented 
with $i_1 = i_2 = 60\degr $ 
and $\omega_1 = \omega_2 = -30\degr $ 
\citep[see][]{1972ApJ...178..623T, 1988ApJ...331..699B}. 
The model galaxies are initialized near the apocentre of 
the corresponding elliptical Keplerian orbit which is defined by the 
eccentricity 
of $e = 0.5$, the 
pericentric distance of $r_{\mathrm{peri}}=20$~kpc, 
and the period of 
1.2~Gyr. The time step in the
simulations is fixed to 1~Myr which guarantees 
that energy is conserved to 
better than 1\% during the merger.
The true decaying orbit is characterized by the generalized 
eccentricity of
$e = 0.8$ and the pericentric distance of 
$r_{\mathrm{peri}}=7$~kpc.
The outer regions of the 
two galaxy bulges begin to merge 
after $\sim 272.5$~Myr, just before pericentric passage. 
The merging is not a single process but the centres of the two bulges
are repeatedly flung apart before they settle down in one common density
peak.
  
\begin{figure*}
        \centering\resizebox{\hsize}{!}{\includegraphics{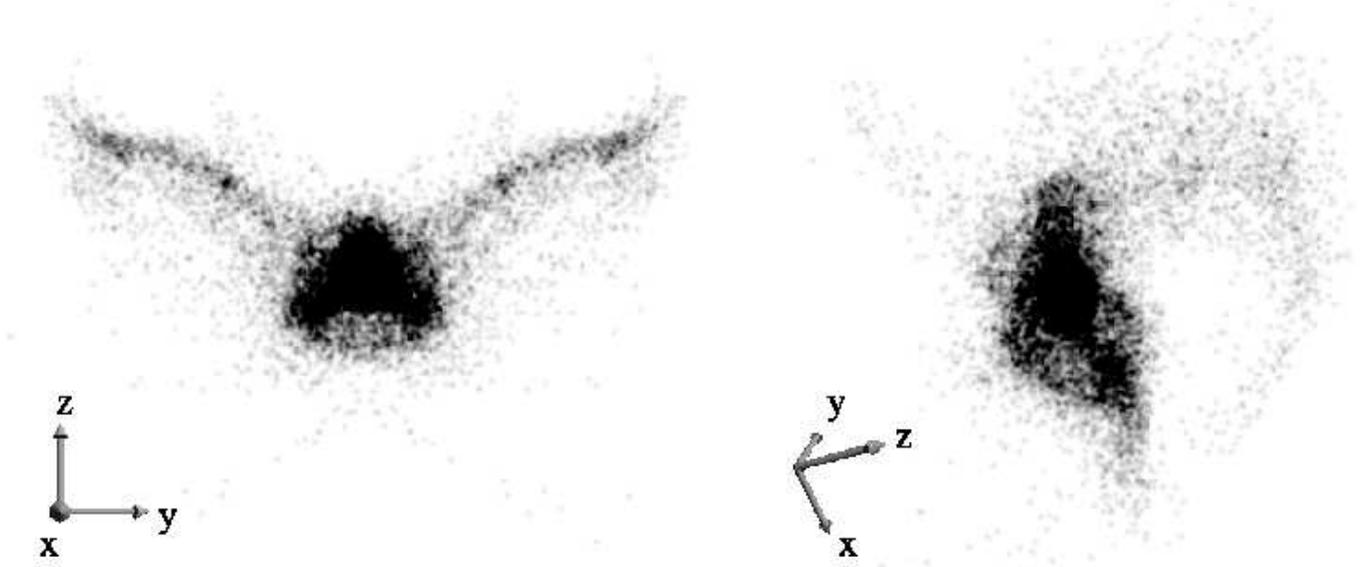}}
        \caption{Two different projections of the same simulation 
	snapshot after 461.1~Myr. The left panel shows the projection 
	for which the merger remnant looks like the ''Antennae'',
	the right panel shows the projection for which the remnant
	looks like the \object{3C 48} host.
        The coordinate planes indicate the respective tilt of the orbital
        plane (x-y). 
        See text for a detailed description.}
        \label{fig:antennae}
\end{figure*} 

\begin{figure}
        \centering\resizebox{\hsize}{!}{\includegraphics{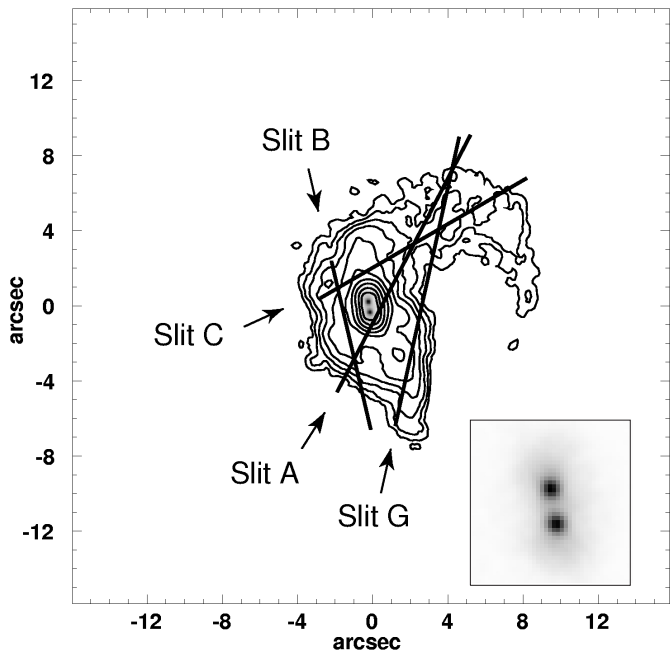}}
        \caption{Contour plot of the surface brightness of the 
        \object{3C 48} look-alike. Length units are fixed in arcsec by 
        positioning the four slits A, B, C, G used by 
        \citet{2000ApJ...528..201C}.
        The small inset shows a magnified view on the still separated 
	density peaks of the merging bulges.}
        \label{fig:simdens}
\end{figure}

\begin{figure*}
        \centering\resizebox{\hsize}{!}{\includegraphics{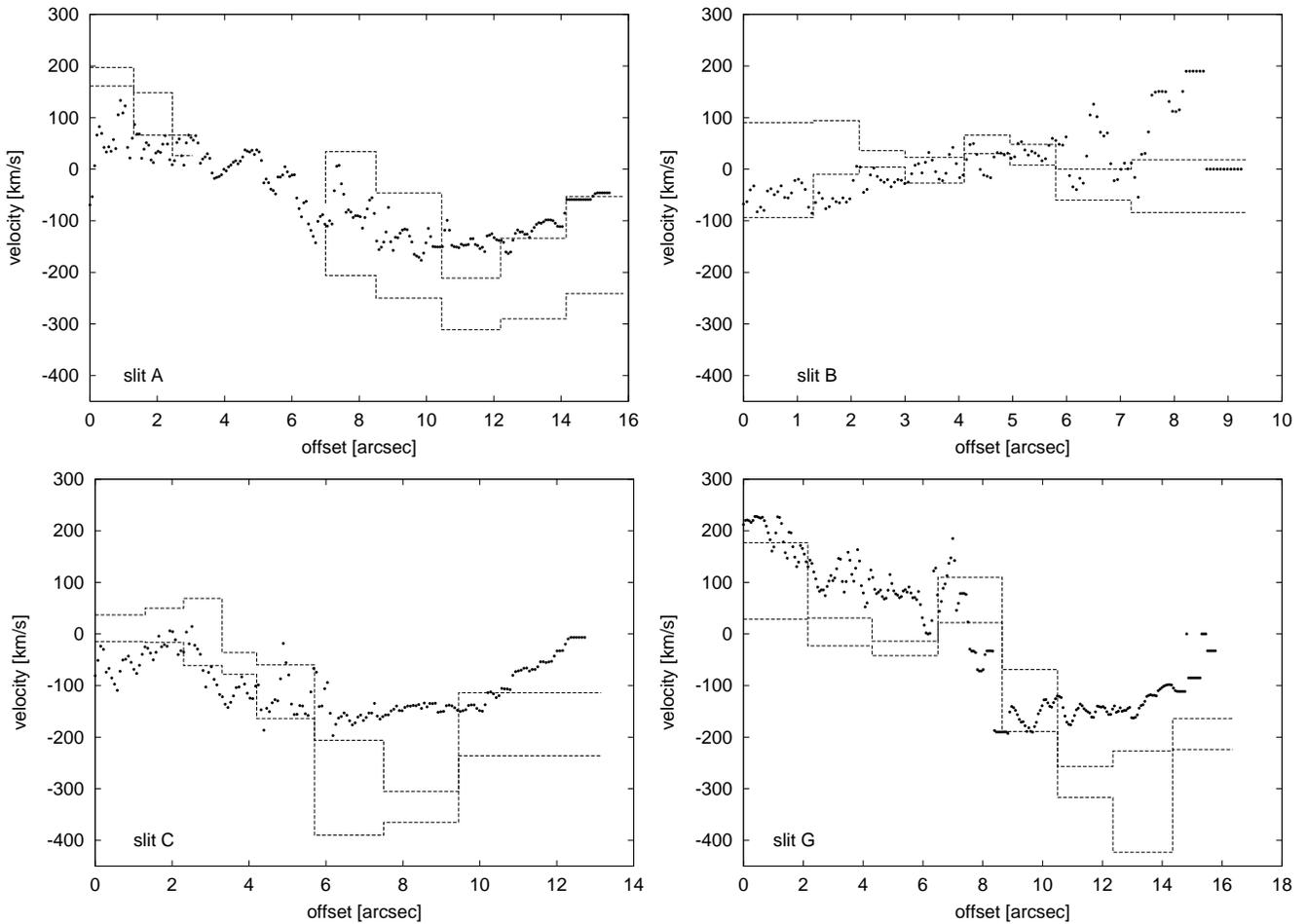}}
        \caption{Comparison of the observed and simulated 
          line-of-sight velocities $v_\mathrm{LOS}$ 
          along the four slits A, B, C, G.
	  Offsets are measured in the direction indicated by the arrows
	  in Fig.~\ref{fig:simdens}. 
          Dots
          represent the average velocities of the \object{3C 48} look-alike.
          The area enclosed by $v_\mathrm{LOS}+\Delta v_\mathrm{LOS}$ and 
          $v_\mathrm{LOS}-\Delta v_\mathrm{LOS}$ (dashed histograms)      
          corresponds to the          
          confidence region of
          the stellar line-of-sight velocities measured for the 
          \object{3C 48} host. These data 
	are taken from Fig.~1 and Table~2 in
         \citet{2000ApJ...528..201C}.}
        \label{fig:vlosfit}
\end{figure*}

\begin{figure}
        \centering\resizebox{\hsize}{!}{\includegraphics{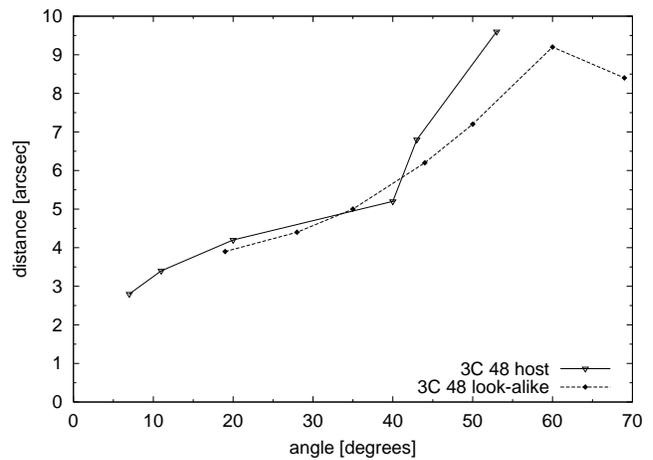}}
        \caption{Comparison of the curvature of the northwestern tidal tail of
        the \object{3C 48} host (solid line) and its look-alike (dashed line).}
        \label{fig:curv}
\end{figure}

\begin{table}
        \caption{Measured proportions of the \object{3C 48} host, as taken 
        from the
        contour plot in Fig.~1 of \citet{2000ApJ...528..201C}, compared to the
        proportions of the nearest look-alike.}
        \label{tab:ratios}
        $$
        \begin{array}{l|cc}
        \hline
        \noalign{\smallskip}
        \hline
        \mathrm{Ratios} & \mathrm{\object{3C\ 48}\ host} & \mathrm{Look-alike} \\ 
        \hline
        L1/L2 & 0.76 & 0.70 \\
        L1/W1 & 1.25 & 1.44 \\
        L1/W2 & 1.67 & 1.86 \\
        L1/D   & 0.52 & 0.55 \\
        L2/W1 & 1.65 & 2.06 \\
        L2/W2 & 2.20 & 2.64 \\
        L2/D   & 0.69 & 0.79 \\
        W1/W2 & 1.33 & 1.29 \\
        W1/D   & 0.42 & 0.38 \\
        W2/D   & 0.31 & 0.30 \\ 
        \hline
        \end{array}
        $$
\end{table} 
The nearest \object{3C 48} look-alike emerges after 461.1~Myr. 
Two projections of this merger remnant are shown in Fig.~\ref{fig:antennae}. 
In the left panel (''Antennae'' look-alike), the view is perpendicular to 
the orbital plane x-y. In the right panel, 
(\object{3C 48} look-alike) the orbital 
plane is tilted southwards, 
westwards, and counterclockwise by $120\degr $, $160\degr $, 
and $116\degr $, respectively.  
The proportions of the \object{3C 48} look-alike are listed in the 
right column of 
Table~\ref{tab:ratios}. 
The positions of the four slits A, B, C, G 
on the look-alike host and the resulting 
physical coordinate system are shown in Fig.~\ref{fig:simdens}. 
In Fig.~\ref{fig:vlosfit}, the scaled velocities along the 
slits are compared to the 
confidence region of stellar line-of-sight
velocities given for \object{3C 48} by \citet{2000ApJ...528..201C}.
The angle-versus-distance comparison for the 
curvature of the northwestern tidal tail is shown in 
Fig.~\ref{fig:curv}.

\section{Discussion\label{sec:discussion}}

\subsection{The nature of \object{3C 48A}}

Optical and near-infrared images of the \object{3C 48} host show 
two luminosity peaks at the positions of \object{3C 48} and \object{3C 48A},
the latter being located about
$1\arcsec $ northeast of \object{3C 48} \citep[e.g.][]{1991AJ....102..488S,
2000ApJ...528..201C}.
With softening lengths of $\sim 0.25$~kpc ($0\farcs06 $)
for the bulge and disk
particles (see Table~\ref{tab:galparams}), 
the spatial resolution of the simulations is 
high enough to identify corresponding density peaks
in the \object{3C 48} model. 
As shown in the small inset in Fig.~\ref{fig:simdens}, the centres of the
bulge components of the two merging galaxies are still separated at the 
stage of the \object{3C 48} look-alike.  
Their distance of about $0\farcs 6 $ (2.5~kpc) and their
relative positions on a southwestern to northeastern axis are 
similar to the observed configuration of \object{3C 48} and 
\object{3C 48A}. Thus, a scenario with 
\object{3C 48} and \object{3C 48A} being the two centres of 
merging galaxies is possible.   
However, the exact configuration of the density peaks in the simulations 
is very sensitive to the projection angle 
and to the time at which the snapshot is taken. About 20~Myr 
later, the two 
peaks have already merged into one. Furthermore, the purely 
stellar-dynamical model does not address the question of a possible
nuclear activity at the positions of \object{3C 48}
and \object{3C 48A}. A detailed discussion of nuclear activity depends 
on whether or not a black hole exists at the mentioned positions and on the
respective fueling rates.

\subsection{Location of the counter tidal tail}
Since each of the two merging disk galaxies 
forms a tidal tail, the missing second tidal tail has always been
a caveat of the merger hypothesis for \object{3C 48}. Here, the simulations
suggest a simple solution: 
At the projection angle of the \object{3C 48}
look-alike, the second tidal tail is mainly located in front of the
body of the host and, therefore, severely foreshortened. It extends from
the southwest towards the northeast, roughly along slit B, so that 
measurements along this slit trace a mixture of 
line-of-sight velocities from the body and from the tail. 
This could explain why the observed and the simulated 
line-of-sight velocity signatures along slit B (Fig.~\ref{fig:vlosfit})
are dominated 
by scattering around a mean velocity close to zero. 
Slits A, C, and G, 
in contrast, are characterized by large absolute 
line-of-sight velocities (up to $\mathrm{\sim 200\ to\ 300\ km\ s^{-1}}$)
and strong variations along the slits.
A counter tidal tail extending from the southwest towards the northeast 
in front of the main body of the \object{3C 48} host is a completely new
alternative. Regarding the information about stellar kinematics, 
this location seems to be more likely than the two suggested tails
in the southeast and from the southeast towards the southwest 
\citep{1999MNRAS.302L..39B, 2000ApJ...528..201C} which have failed 
identification so far.

\subsection{The evolutionary history of \object{3C 48}}

Conclusions about the orbital parameters for \object{3C 48} and the 
original parameters of the merging galaxies can only be tentative.
The orbital period of the best fit model amounts to about
20\% of the
age of the universe at the redshift of \object{3C 48} 
($\sim 5.4$~Gyr).
A merger scenario with such an orbital period 
is plausible, assuming an initially highly eccentric orbit of the 
merging galaxies which is transformed into a bound orbit by dynamical 
friction of their dark matter halos \citep[e.g.][]{1989AJ.....98.1557J}.
It has been found 
that the morphology and the kinematics of tidal tails are very sensitive to
the rotation curve of the interacting model galaxies
\citep{1996ApJ...462..576D, 1998ApJ...494..183M, 1999ApJ...526..607D}. 
Thus, instead of two identical galaxies, an alternative 
model for \object{3C 48} could start from two galaxies with different
rotation curves so that only one of them forms an extended tidal tail.
However, even in its generality 
the multi-particle model presented in this paper gives rather 
compelling
evidence that the formation of \object{3C 48} is linked to a merger
process. 
Therewith, \object{3C 48} ranks among these transitional objects which 
support the evolutionary 
scenario \citep{1988ApJ...325...74S} in its original merger-driven 
definition.

\bibliographystyle{aa}
\bibliography{h4735.bib}

\begin{thebibliography}{22}
\expandafter\ifx\csname natexlab\endcsname\relax\def\natexlab#1{#1}\fi

\bibitem[{{Barkhouse} \& {Hall}(2001)}]{2001AJ....121.2843B}
{Barkhouse}, W.~A. \& {Hall}, P.~B. 2001, \aj, 121, 2843

\bibitem[{{Barnes}(1988)}]{1988ApJ...331..699B}
{Barnes}, J.~E. 1988, \apj, 331, 699

\bibitem[{{Boroson} \& {Oke}(1982)}]{1982Natur.296..397B}
{Boroson}, T.~A. \& {Oke}, J.~B. 1982, \nat, 296, 397

\bibitem[{{Boyce} {et~al.}(1999){Boyce}, {Disney}, \&
  {Bleaken}}]{1999MNRAS.302L..39B}
{Boyce}, P.~J., {Disney}, M.~J., \& {Bleaken}, D.~G. 1999, \mnras, 302, L39

\bibitem[{{Canalizo} \& {Stockton}(2000)}]{2000ApJ...528..201C}
{Canalizo}, G. \& {Stockton}, A. 2000, \apj, 528, 201

\bibitem[{{Chatzichristou} {et~al.}(1999){Chatzichristou}, {Vanderriest}, \&
  {Jaffe}}]{1999A&A...343..407C}
{Chatzichristou}, E.~T., {Vanderriest}, C., \& {Jaffe}, W. 1999, \aap, 343, 407

\bibitem[{{Dubinski} {et~al.}(1996){Dubinski}, {Mihos}, \&
  {Hernquist}}]{1996ApJ...462..576D}
{Dubinski}, J., {Mihos}, J.~C., \& {Hernquist}, L. 1996, \apj, 462, 576

\bibitem[{{Dubinski} {et~al.}(1999){Dubinski}, {Mihos}, \&
  {Hernquist}}]{1999ApJ...526..607D}
---. 1999, \apj, 526, 607

\bibitem[{{Hernquist}(1993)}]{1993ApJS...86..389H}
{Hernquist}, L. 1993, \apjs, 86, 389

\bibitem[{{Hernquist} \& {Katz}(1989)}]{1989ApJS...70..419H}
{Hernquist}, L. \& {Katz}, N. 1989, \apjs, 70, 419

\bibitem[{{Jones} \& {Stein}(1989)}]{1989AJ.....98.1557J}
{Jones}, B. \& {Stein}, W.~A. 1989, \aj, 98, 1557

\bibitem[{{Ma} {et~al.}(1998){Ma}, {Arias}, {Eubanks}, {Fey}, {Gontier},
  {Jacobs}, {Sovers}, {Archinal}, \& {Charlot}}]{1998AJ....116..516M}
{Ma}, C., {Arias}, E.~F., {Eubanks}, T.~M., {et~al.} 1998, \aj, 116, 516

\bibitem[{{Matthews} {et~al.}(1961){Matthews}, {Bolton}, {Greenstein}, {Munch},
  \& {Sandage}}]{1961ST....21.148M}
{Matthews}, T.~A., {Bolton}, J.~G., {Greenstein}, J.~L., {Munch}, G., \&
  {Sandage}, A.~R. 1961, Sky and Telescope, 21, 148

\bibitem[{{Matthews} \& {Sandage}(1963)}]{1963ApJ...138...30M}
{Matthews}, T.~A. \& {Sandage}, A.~R. 1963, \apj, 138, 30

\bibitem[{{Mihos} {et~al.}(1998){Mihos}, {Dubinski}, \&
  {Hernquist}}]{1998ApJ...494..183M}
{Mihos}, J.~C., {Dubinski}, J., \& {Hernquist}, L. 1998, \apj, 494, 183

\bibitem[{{Neugebauer} {et~al.}(1985){Neugebauer}, {Soifer}, \&
  {Miley}}]{1985ApJ...295L..27N}
{Neugebauer}, G., {Soifer}, B.~T., \& {Miley}, G.~K. 1985, \apjl, 295, L27

\bibitem[{{Sanders} {et~al.}(1988){Sanders}, {Soifer}, {Elias}, {Madore},
  {Matthews}, {Neugebauer}, \& {Scoville}}]{1988ApJ...325...74S}
{Sanders}, D.~B., {Soifer}, B.~T., {Elias}, J.~H., {et~al.} 1988, \apj, 325, 74

\bibitem[{{Scoville} {et~al.}(1993){Scoville}, {Padin}, {Soifer}, {Yun}, \&
  {Sanders}}]{1993AAS...182.0624S}
{Scoville}, N.~Z., {Padin}, S., {Soifer}, B.~T., {Yun}, M.~S., \& {Sanders},
  D.~B. 1993, Bulletin of the American Astronomical Society, 25, 1241

\bibitem[{{Stockton} \& {Ridgway}(1991)}]{1991AJ....102..488S}
{Stockton}, A. \& {Ridgway}, S.~E. 1991, \aj, 102, 488

\bibitem[{{Toomre} \& {Toomre}(1972)}]{1972ApJ...178..623T}
{Toomre}, A. \& {Toomre}, J. 1972, \apj, 178, 623

\bibitem[{{Wilkinson} {et~al.}(1991){Wilkinson}, {Tzioumis}, {Benson},
  {Walker}, {Simon}, \& {Kahn}}]{1991Natur.352..313W}
{Wilkinson}, P.~N., {Tzioumis}, A.~K., {Benson}, J.~M., {et~al.} 1991, \nat,
  352, 313

\bibitem[{{Wink} {et~al.}(1997){Wink}, {Guilloteau}, \&
  {Wilson}}]{1997A&A...322..427W}
{Wink}, J.~E., {Guilloteau}, S., \& {Wilson}, T.~L. 1997, \aap, 322, 427

\end{thebibliography}

\begin{acknowledgements}
Our special thanks go to Prof. Dr Lars Hernquist who kindly provided the
codes TREESPH and BUILDGAL and gave helpful advice.
This project was supported in part 
by the Deutsche Forschungsgemeinschaft (DFG) via
grant SFB 494.
J. Scharw\"achter is supported by a scholarship for doctoral students of the
Studienstiftung des deutschen Volkes. 
\end{acknowledgements}

\end{document}